\documentclass[twocolumn,showpacs,aps,epsfig]{revtex4}

\usepackage{epsfig}
\usepackage{graphicx}
\usepackage{dcolumn}
\usepackage{bm}
\usepackage{epstopdf}
\begin{document}
\newcommand{\be}{\begin{equation}}
\newcommand{\ee}{\end{equation}}
\newcommand{\bea}{\begin{eqnarray}}
\newcommand{\eea}{\end{eqnarray}}
\title{\bf Transport and Helfand moments in the Lennard-Jones fluid. II. Thermal conductivity}
\author{S. Viscardy, J. Servantie, and P. Gaspard\\
{\em Center for Nonlinear Phenomena and Complex Systems,}\\
{\em Universit\'e Libre de Bruxelles,}\\
{\em Campus Plaine, Code Postal 231, B-1050 Brussels, Belgium}\\}
\begin{abstract}
The thermal conductivity is calculated with the Helfand-moment method
in the Lennard-Jones fluid near the triple point. 
The Helfand moment of thermal conductivity 
is here derived for molecular dynamics with periodic boundary conditions.
Thermal conductivity is given by a generalized Einstein relation
with this Helfand moment. We compute thermal conductivity by this new
method and compare it with our own values obtained by the standard
Green-Kubo method. The agreement is excellent.
\end{abstract}
\maketitle
\section{Introduction}
This paper completes the analysis carried out in the companion paper \cite{VSG06-1} by the study of thermal conductivity with the Helfand-moment method \cite{helf}.
Since the seventies, several works have been carried out for the calculation of the transport coefficients. Shear viscosity is the most studied transport property, while thermal conductivity has
been less studied. The first computation of thermal conductivity goes back 
to the work of Alder \textit{et al.} for hard-sphere systems \cite{alder}. Soft-sphere potential systems were considered for the first time a few years later by
Levesque \textit{et al.} \cite{LVK-73} using the well-known Green-Kubo formula \cite{green51,green60,kubo57,mori58}.
Nevertheless, since the autocorrelation function of the microscopic flux decreases slowly, especially near the triple point, most of the studies devoted to the computation of thermal conductivity were performed by nonequilibrium methods. This is usually done by fixing the temperature gradient and measuring the heat flux \cite{ashurst73,evans82,ciccotti-tenenbaum80,MC84,evans86,heyes-88}, except in Ref. \cite{muller-plathe97} where the opposite is done, i.e., the heat flux is fixed and the temperature gradient is measured. However, some equilibrium molecular dynamics studies using the standard Green-kubo formula were carried out \cite{LVK-73,MC84,heyes-88,hoheisel-90}. More recently, as suggested in the nineties \cite{allen93,haile}, the generalized Einstein relation expressing thermal conductivity as the variance of the time integral of the microscopic flux was computed \cite{meier-thesis}.
The purpose of the present paper is to extend the Helfand-moment method of the companion paper \cite{VSG06-1} from viscosity to thermal conductivity.  As an application of our method, we calculate the thermal conductivity of the Lennard-Jones fluid at a phase point near the triple point. We show that the values obtained by the Helfand-moment method and the Green-Kubo values are in excellent agreement.
Such a method brings not only some interest as an alternative equilibrium molecular dynamics method, but also in the context of
recent theories in nonequilibrium statistical mechanics. Indeed, the {\it escape-rate formalism} propose clear and direct relationships between the transport coefficients and quantities characterizing the microscopic chaos
such as the Lyapunov exponents and fractal dimensions \cite{dorf-gasp,gasp-dorf,gasp-book,dorf-book}. Successful results have already been obtained for simple systems: diffusion in the Lorentz gas \cite{gasp-baras} and viscosity in a two-particle system \cite{viscardy-gaspard2}. Furthermore, the {\it hydrodynamic-mode method}, aiming to construct at the microscopic level the hydrodynamic modes which are the solutions of
the diffusion or Navier-Stokes equations also depend on the possibility to define Helfand moments in periodic systems. This construction has already been carried out  for diffusion in the Lorentz gas \cite{gaspard96}. By defining the Helfand moment, we hopefully project to extend this construction to the other transport processes, in particular thermal conductivity.
The paper is organized as follows. In Section \ref{TB}, the theoretical results of Helfand \cite{helf} are briefly outlined. In Section \ref{HM}, we present our {\it Helfand-moment method} for the calculation of thermal conductivity.  Section \ref{Num} gives the results of the molecular dynamics simulations. We compare our results by the Heldand-moment and Green-Kubo techniques and also 
to the litterature. Finally, conclusions are drawn in Section \ref{Conclusions}.
\section{Theoretical background}
\label{TB}
In the fifties, the Boltzmann and Enskog theories of transport processes were completed by an alternative approach developed by Green \cite{green51,green60}, Kubo \cite{kubo57} and Mori \cite{mori58}.  This new approach relates the transport
coefficients to the time autocorrelation functions of their corresponding microscopic flux. In particular, the thermal conductivity
coefficient $\kappa$ can be expressed as
\begin{equation}
\kappa = \lim_{N,V\to\infty} \frac{1}{k_{\rm B} T^2V}\int_{0}^{\infty}  \left \langle J^{(\kappa)}(0) J^{(\kappa)}(t)\right \rangle \; dt \, .
\label{GK.thermal.conductivity}
\end{equation}
where $J^{(\kappa)}$ is the microscopic flux associated with thermal conductivity.
This theory plays an important role in numerical simulations. On the other hand, Helfand aimed to express the transport coefficients in terms of generalized Einstein relations \cite{helf} and showed that thermal conductivity is given by
\begin{equation}
\kappa=\lim_{N,V,t \to \infty}\frac{1}{2k_{\rm B} T^2Vt}  \left \langle \left[G^{(\kappa)} (t)-G^{(\kappa)} (0) \right]^{2}\right \rangle , 
\label{Einstein.thermal.conductivity}
\end{equation}
where $G^{(\kappa)}(t)$ is the corresponding Helfand moment defined as the centroid \begin{equation}
G^{(\kappa)} (t) = \sum_{a=1}^{N} x_{a} \left ( E_a - \left
\langle E_a \right \rangle \right ) \, ,
\label{thermal-conductivity-helfand-moment}
\end{equation}
of the energies of the particles
\begin{equation}
E_a \equiv \frac{{\bf p}_a^2}{2m} + \frac{1}{2} \sum_{b(\ne a)} u({\bf r}_{ab}) \, .
\end{equation}
It can be proved that the Green-Kubo formula (\ref{GK.thermal.conductivity}) and the Helfand equation (\ref{Einstein.thermal.conductivity}) are equivalent if the microscopic flux is related to the Helfand moment by
\begin{equation}
J^{(\kappa)} (t) = \frac{d G^{(\kappa)} (t) }{dt} \; .
\end{equation}
Thanks to Eq. (\ref{Einstein.thermal.conductivity}), the thermal conductivity coefficient is manifestly a non-negative quantity, as required by the second law of thermodynamics.
\section{Helfand-moment method}
\label{HM}
Since the transport coefficients are bulk properties, they can be calculated with
equilibrium molecular dynamics with periodic boundary conditions.  This dynamics is ruled by Newton's equations modified in order to take into account the periodicity:
\begin{eqnarray}
\frac{d{\bf r}_a}{dt} &=& \frac{{\bf p}_a}{m} + \sum_{s}
\Delta{\bf r}_{a}^{(s)} \; \delta(t-t_{s})\; , \nonumber \\
\frac{d{\bf p}_{a}}{dt} &=& \sum_{b (\neq a)} {\bf F}({\bf
r}_{ab}) \; ; \qquad {\bf F}({\bf r}) = -
\frac{\partial u({\bf r})}{\partial {\bf r}} \; ,
\label{Newton}
\end{eqnarray}
where $\Delta{\bf r}_{a}^{(s)}$ is the jump of the particle $a$ at time $t_{s}$ with $\Vert \Delta {\bf r}_{a}^{(s)}  \Vert = L$ equal to the length $L$ of the simulation box. The force ${\bf F}({\bf r})$ is here assumed of finite range $r<L/2$.  The relative position appearing in the force is defined as
\begin{equation}
{\bf r}_{ab} = {\bf r}_a - {\bf r}_b - {\bf L}_{a|b}
\end{equation}
where the time-dependent cell translation vector ${\bf L}_{a|b}$ \cite{haile} is chosen for the minimum-image convention $\Vert {\bf r}_{ab} \Vert < L / 2$ to be satisfied.  For more details, see Section III of the companion paper \cite{VSG06-1}.
We propose here a Helfand-moment method for thermal conductivity in systems with periodic boundary conditions. As for shear viscosity \cite{VSG06-1}, Helfand's original expression (\ref{thermal-conductivity-helfand-moment}) must be modified in order to take into account the periodicity of the system by the addition of extra terms:
\begin{equation}
G^{(\kappa)} (t) = \sum_{a=1}^{N} x_{a} \left ( E_a - \left \langle E_a \right \rangle \right ) + I(t) \, .
\label{thermal-conductivity-helfand-moment-pbc}
\end{equation}
Following a way similar to the case of viscosity, we conclude that the Helfand moment which should be used in
periodic systems is given by
\begin{eqnarray}
G^{(\kappa)} (t) & = & \sum_{a}^{} x_{a} \left ( E_a - \left
\langle E_a \right \rangle \right ) \nonumber \\ & - & \sum_{a}^{}
\sum_{s} \Delta x_{a}^{(s)} \left ( E_a - \left \langle E_a \right
\rangle \right ) \theta(t -t_s) \nonumber \\ & - & \frac{1}{4}
\sum_{a \ne b}^{} \int_{0}^{t} d \tau \;  L_{b|a x} \frac{{\bf
p}_a + {\bf p}_b}{m} \cdot {\bf F}({\bf r}_{ab})\, .
\label{heat-conductivity-helfand}
\end{eqnarray}
The derivation is given in Appendix A. As for shear viscosity, the first extra term is due to the jumps of particles from a boundary to the opposite one. The second extra term takes into account of interactions of particles with the image of other particles. By adding the two extra terms to the original Helfand moment, we recover the relation $G=\int dt J$. 
\section{Numerical results}
\label{Num}
\begin{figure}[h!]
\includegraphics[scale=0.6]{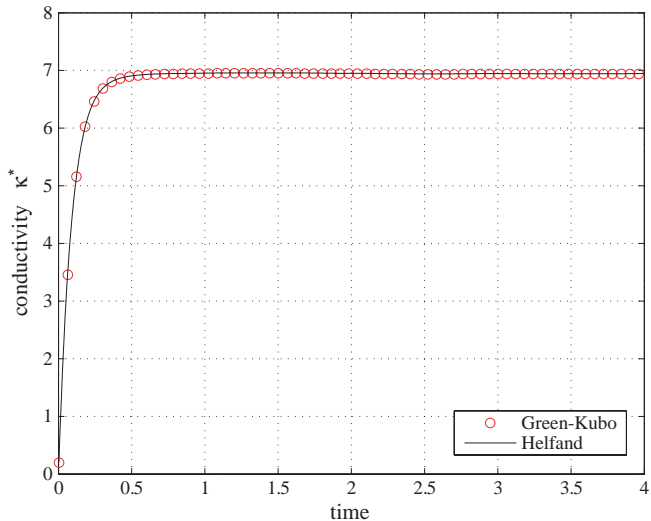}
\caption{Thermal conductivity at the phase point $T^*=0.722$ and
$n^*=0.8442$ for $N=1372$. The plain line is the derivative of
the mean-square displacement of the Helfand moment and the circles
the integral of the microscopic flux autocorrelation function.}
\label{kappacomp}
\end{figure}

\begin{figure}[h!]
\includegraphics[scale=0.6]{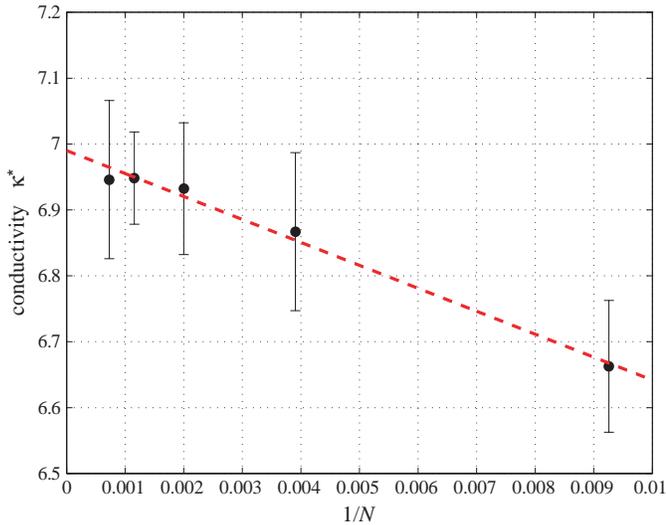}
\caption{Thermal conductivity at the phase point $T^*=0.722$ and
$n^*=0.8442$ as a function of the inverse of the number $N$ of atoms. The
circles are the results of the numerical simulations and the
dashed line the linear extrapolation.} 
\label{kappaconv}
\end{figure}

We carried out molecular dynamics simulation to apply our Helfand-moment method to the calculation of thermal conductivity in a fluid with the standard 6-12
Lennard-Jones potential $u(r)=4\epsilon\left[(\sigma/r)^{12}-(\sigma/r)^{6}\right]$. All calculations we performed are done with the cutoff  $r_c=2.5 \sigma$.  In the reduced units with the time $t^* = t\sqrt{\epsilon/(m \sigma^2)}$ and the space $r^*= r/\sigma$, the reduced thermal conductivity is given by
\begin{equation}
\kappa^* = \kappa\,\frac{\sigma^2}{k_{\rm B}} \sqrt{\frac{m}{\epsilon}} \, .
\end{equation}
The details of the simulations are given in the preceding paper \cite{VSG06-1}. To show the validity of the Helfand-moment method, we first 
compare with the results of the Green-Kubo formula. We depict in Fig. \ref{kappacomp} the time derivative of the
mean-square displacement of the Helfand moment (\ref{thermal-conductivity-helfand-moment}) and the time integral of the autocorrelation function of the microscopic flux. The calculation is carried out for $N=1372$ atoms at the phase point near of the triple point with a reduced temperature of $T^*=k_{\rm B}T/\epsilon=0.722$ and a density $n^*=n\sigma^3=0.8442$. As seen in Fig. \ref{kappacomp}, the two methods are in perfect agreement.
We estimated thermal conductivity by a linear fit on the mean-square displacement of the Helfand moment. The fit is done in the region between 2 and 8 time units to guarantee that the linear regime is reached.  We calculated the thermal conductivity for increasing system sizes from $N=108$ atoms to 
$N=1372$. We give the results in Table \ref{TABCOMP} and depict them in Fig. \ref{kappaconv} as a function of the inverse $N^{-1}$ of the system size.  The linear extrapolation gives the following estimate of thermal conductivity for an infinite system,
\begin{equation}
\kappa^*=6.990 \pm 0.030 \; .
\end{equation}
We see in Table \ref{TABCOMP} that the values of the present work compare very well with the values of previous studies. 
\begin{table}[n]
\begin{center}
\begin{tabular}{lccccccc}
\hline 
\hline
Authors & year & method & $r^{*}_{\rm cut}$	 & $T^*$ & $N$ & $\kappa^*$ 	& $\Delta \kappa^*$\\ 
\hline 
Levesque \textit{et al.} \cite{LVK-73}  & 1973 & GK & N. C. & 0.715 & 256   & 7.067 	& 0.416\\ 
  &  &  &  & 0.722 & 864  & 7.136 	& 0.277 \\ 
&  & & &       	&   &   &   \\
Evans \cite{evans82} & 1982 & EM & 2.5	& 0.722 & 108 	& 6.54  & N. C. \\ 
&  & & &       	&   &   &   \\
Massobrio and Ciccotti \cite{MC84}      & 1984 & GK  & 3.5 & 0.721 	& 256  & 6.880 	& 0.485\\ 
 &  & DM  & 3.5	    & 0.721  & 256  & 6.873 	& 0.229\\ 
&  & & &       	&   &   &   \\
Paolini \textit{et al.} \cite{PCM86}    & 1986 & NEM & 2.5	& 0.718-0.721 	& average$^{a)}$ & 6.776 	& 0.166\\ 
&  & & &       	&   &   &   \\
Heyes  \cite{heyes-88} 		& 1988 & GK & N. C.	& 0.72 	& 108  & 6.7    & 0.34 ($5\%$) \\ 
 &  & & & 		& 256 		 & 6.9 		& 0.35 ($5\%$)  \\ 
 &  & & &       	& 500 		& 6.5 		& 0.33 ($5\%$) \\
&  & & &       	&   &   &   \\
This work & 2007 & HM (MD) & 2.5 & 0.722 & 108 & 6.663  &  0.10 \\ 
 &  & & &       	& 256 & 6.867 & 0.12 \\ 
 &  & & &       	& 500 & 6.932 & 0.10 \\
 &  & & &       	& 864 & 6.948 & 0.07 \\ 
 &  & & &       	& 1372 & 6.946 & 0.12 \\
 &  & & &       	& $\infty$ & 6.990 & 0.061 \\
\hline
\hline
\end{tabular}
\caption{Results found in the literature for the thermal
conductivity in the Lennard-Jones fluid near the triple point. The
reduced density is $n^* = 0.8442$ except for Heyes \cite{heyes-88}
($n^*=0.848)$.
Abbreviations: DM, nonequilibrium results obtained by the
differential method. EM, Evans method \cite{evans82}. GK,
Green-Kubo formula with equilibrium molecular dynamics. NEM,
results by nonequilibrium method obtained by Paolini, Ciccotti and
Massobrio \cite{PCM86}. HM, the present Helfand-moment method with molecular dynamics (MD).
N.~C. means the value has not been communicated.\\
$^{a)}$ Thermal conductivity obtained by averaging the values for $N=108, 256, 500, 864$.}
\label{TABCOMP} 
\end{center}
\end{table}
\section{Conclusions}
\label{Conclusions}
In this paper, a new method is proposed for the calculation of the thermal conductivity in soft-sphere systems. The technique we call the \textit{Helfand-moment method} is based on generalized Einstein relations (\ref{Einstein.thermal.conductivity}) with the Helfand moment defined to take into account
periodic boundary conditions in the molecular dynamics.
This allows us to calculate the thermal conductivity coefficient in terms of the variance of the Helfand moment (\ref{heat-conductivity-helfand}). We show that the original Helfand moment (\ref{thermal-conductivity-helfand-moment}) must be modified by adding two extra terms. After the addition of these extra terms, we recover the relation $J=\dot{G}$ between the microscopic flux $J$ and the corresponding Helfand moment $G$. This is the generalization of the case of diffusion for which the velocity $v_x$ (the microscopic flux) is related to the position $x$ (the Helfand moment) in the same way $v_x=\dot{x}$. Consequently, we have showed that it is possible to calculate the thermal conductivity coefficient by the mean-square displacement of the Helfand moment. Indeed, our molecular dynamics simulations show perfect agreement between the Helfand-moment and Green-Kubo methods. We think that the new expressions for the Helfand moment given in the present and the companion papers \cite{VSG06-1} solve the problems on the use of generalized Einstein relations in periodic systems. 
In addition to the viscosities \cite{VSG06-1} and thermal conductivity, it is possible to use a similar method for the
electric conductivity $\sigma$ in periodic systems. Indeed, by a similar derivation to the one found in Appendix A, one can show that the
electric conductivity can be written as:
\begin{equation}
\sigma = \lim_{N,V,t \to \infty}  \frac{1}{2k_{\rm B} TVt}  \left\langle \left[ G^{(\sigma)}(t) - \langle G^{(\sigma)}(t) \rangle \right]^2 \right\rangle ,
\end{equation}
where the modified Helfand moment for periodic
systems is expressed as:
\begin{equation}
G^{(\sigma)}(t) = \sum_{a} e Z_a \left[ x_a - \sum_s \Delta
x_a^{(s)} \theta (t -t_s) \right] ,
\end{equation}
with the electric charge $eZ_a$ of the particles and $G^{(\sigma)}(0)=0$.
Finally, we notice that the Helfand-moment method plays an important role in
the escape-rate formalism \cite{dorf-gasp,gasp-dorf} and the
hydrodynamic-mode method \cite{gaspard96} developed in
nonequibrium statistical mechanics \cite{gasp-book}. They
establish relationships between microscopic and macroscopic levels
in the context of the understanding of the origin of the
irreversibility of chemical-physical phenomena. The derivation of
the Helfand-moment methods give one the possibility to confirm
numerically the theoretical predictions of these formalisms.
\acknowledgments
We thank K. Meier for useful discussions.
This research is financially supported by the ``Communaut\'e fran\c 
caise de Belgique'' (contract ``Actions de Recherche Concert\'ees'' 
No.~04/09-312) and the National Fund for Scientific Research 
(F.~N.~R.~S. Belgium, contract F.~R.~F.~C. No.~2.4577.04).
\appendix
\section{Derivation of the Helfand moment for the thermal conductivity in periodic systems}
The time derivative of the modified Helfand moment (\ref{thermal-conductivity-helfand-moment-pbc}) is given by
\begin{eqnarray}
\frac{dG^{(\kappa)}(t)}{dt} & = & \sum_{a}^{} \frac{d x_{a}}{dt}
\left ( E_a - \left \langle E_a \right \rangle \right ) \nonumber
\\ & + & \sum_{a}^{} x_{a} \frac{d E_{a}}{dt} +
\frac{d I(t)}{dt} \label{Helf-time-derivative}
\end{eqnarray}
where the time derivative of $E_a$ is 
\begin{eqnarray}
\frac{d E_a}{dt} & = & \frac{{\bf p}_a}{m} \cdot \frac{d {\bf p}_a}{dt}
+ \frac{1}{2} \sum_{b (\ne a)}\frac{\partial u_{ab}}{\partial {\bf r}_{ab}} \cdot \frac{{\bf p}_a - {\bf p}_b}{m} \nonumber \\
& = & \frac{1}{2} \sum_{b (\ne a)} \frac{{\bf p}_a + {\bf p}_b}{m} \cdot {\bf F}({\bf r}_{ab}) \; .
\end{eqnarray}
We notice that there is here no jump in position to consider because $\frac{d {\bf r}_{ab}}{dt}$ concerns a relative position
${\bf r}_{ab} = {\bf r}_a - {\bf r}_b - {\bf L}_{b|a}$ which satisfies the minimum image convention within the range of the force.
Symmetrizing the second term in Eq. (\ref{Helf-time-derivative}),
we find
\begin{eqnarray}
\sum_{a}^{} x_{a} \frac{d E_{a}}{dt} & = & \frac{1}{4} \sum_{a
\ne b}^{} x_{a} \frac{{\bf p}_a +  {\bf p}_b}{m} \cdot {\bf
F}({\bf r}_{ab}) \nonumber \\ & + & \frac{1}{4} \sum_{a \ne
b}^{} x_{b} \frac{{\bf p}_a  + {\bf p}_b}{m} \cdot {\bf F}({\bf
r}_{ba}) \nonumber \\ & = &  \frac{1}{4} \sum_{a \ne b}^{}
x_{ab} \frac{{\bf p}_a + {\bf p}_b}{m} \cdot {\bf F}({\bf r}_{ab})
\nonumber \\ & + & \frac{1}{4} \sum_{a \ne b} L_{b|a
x} \frac{{\bf p}_a + {\bf p}_b}{m} \cdot {\bf F}({\bf r}_{ab})
\end{eqnarray}
where ${\bf F}({\bf r}_{ab}) = - {\bf F}({\bf r}_{ba})$ according to Newton's third law and because
$x_{ab}=x_a-x_b-L_{b|a x}$.
Substituting this expression in Eq. (\ref{Helf-time-derivative}), where $\frac{d x_{a}}{dt}$ is given by
modified Newton's equations of motion (\ref{Newton}), we get the expression
\begin{eqnarray}
\frac{dG^{(\kappa)}(t)}{dt}  & = & \sum_{a}^{} \frac{p_{ax}}{m}
\left ( E_a - \left \langle E_a \right \rangle \right ) \nonumber
\\ & + & \sum_{a}^{} \sum_{s} \Delta x_{a}^{(s)} \left ( E_a - \left
\langle E_a \right \rangle \right ) \delta(t -t_s) \nonumber \\ &
+ & \frac{1}{4} \sum_{a\ne b}^{} x_{ab} \frac{{\bf p}_a + {\bf
p}_b}{m} \cdot {\bf F}({\bf r}_{ab}) \nonumber \\ & + &
\frac{1}{4} \sum_{a \ne b}^{} L_{b|a x} \frac{{\bf p}_a + {\bf
p}_b}{m} \cdot {\bf F}({\bf r}_{ab}) \nonumber \\ & + & \frac{d
I(t)}{dt} \; .
\end{eqnarray}
On the other hand, it is well known that the microscopic flux $J^{(\kappa)}$ for thermal conductivity is given by
\begin{eqnarray}
J^{(\kappa)} (t) & = & \sum_{a}^{} \frac{p_{ax}}{m} \left ( E_a -
\left \langle E_a \right \rangle \right ) \nonumber \\ & + &
\frac{1}{4} \sum_{a \ne b}^{} x_{ab} \frac{{\bf p}_a + {\bf
p}_b}{m} \cdot {\bf F}({\bf r}_{ab}) \; .
\end{eqnarray}
Since $\frac{dG^{(\kappa)}(t)}{dt} = J^{(\kappa)} (t)$ by definition, we thus obtain that:
\begin{eqnarray}
\frac{d I(t)}{dt} & = & - \sum_{a}^{} \sum_{s} \Delta x_{a}^{(s)}
\left ( E_a - \left \langle E_a \right \rangle \right ) \delta(t
-t_s) \nonumber \\ & - & \frac{1}{4} \sum_{a\ne b}^{} L_{b|a x}
\frac{{\bf p}_a + {\bf p}_b}{m} \cdot {\bf F}({\bf r}_{ab}) \; .
\end{eqnarray}
Finally, the quantity to be added to the usual Helfand moment (\ref{thermal-conductivity-helfand-moment}) is:
\begin{eqnarray}
I(t) & = & - \sum_{a}^{} \sum_{s} \Delta x_{a}^{(s)} \left ( E_a -
\left \langle E_a \right \rangle \right ) \theta(t -t_s) \nonumber
\\ & - & \frac{1}{4} \sum_{a \ne b}^{} \int_{0}^{t} d \tau \;
L_{b|a x} \frac{{\bf p}_a + {\bf p}_b}{m} \cdot {\bf F}({\bf
r}_{ab}) \; .
\end{eqnarray}

\end{document}